\documentclass[11pt,ams]{article}%
\usepackage{geometry}
\usepackage{dsfont}
\usepackage{amsmath}
\usepackage{amsfonts}
\usepackage{amssymb}
\usepackage{graphicx}%
\usepackage{subfig}
\usepackage{float,pslatex}

\newcommand{\p}{\partial}

\begin{document}

\date{}
\title{\textbf{An all-order proof of the equivalence between Gribov's no-pole and Zwanziger's horizon conditions}}
\author{\textbf{M.~A.~L.~Capri}$^{a}$\thanks{caprimarcio@gmail.com}\,\,,
\textbf{D.~Dudal}$^{b}$\thanks{david.dudal@ugent.be}\,\,,
\textbf{M.~S.~Guimaraes}$^{a}$\thanks{msguimaraes@uerj.br}\,\,,
\textbf{L.~F.~Palhares }$^{a,c}$\thanks{leticia@if.ufrj.br}\,\,,\\
\textbf{S.~P.~Sorella}$^{a}$\thanks{sorella@uerj.br}\ \thanks{Work supported by
FAPERJ, Funda{\c{c}}{\~{a}}o de Amparo {\`{a}} Pesquisa do Estado do Rio de
Janeiro, under the program \textit{Cientista do Nosso Estado}, E-26/101.578/2010.}\,\,\\[2mm]
{\small \textnormal{$^{a}$  \it Departamento de F\'{\i }sica Te\'{o}rica, Instituto de F\'{\i }sica, UERJ - Universidade do Estado do Rio de Janeiro,}}
 \\ \small \textnormal{\phantom{$^{a}$} \it Rua S\~{a}o Francisco Xavier 524, 20550-013 Maracan\~{a}, Rio de Janeiro, Brasil}\\
	 \small \textnormal{$^{b}$ \it Ghent University, Department of Physics and Astronomy, Krijgslaan 281-S9, 9000 Gent, Belgium}\\
	 \small \textnormal{$^{c}$ \it Institut f\"ur Theoretische Physik, Heidelberg University, Philosophenweg 16,  69120 Heidelberg, Germany}\normalsize}
\maketitle
\begin{abstract}
\noindent The quantization of non-Abelian gauge theories is known to be plagued by Gribov copies. Typical examples are the copies related to zero modes of the Faddeev-Popov operator, which give rise to singularities in the ghost propagator. In this work we present an exact and compact expression for the ghost propagator as a function of external gauge fields, in $SU(N)$ Yang-Mills theory in the Landau gauge. It is shown, to all orders, that the condition for the ghost propagator not to have a pole, the so-called Gribov's no-pole condition, can be implemented by demanding a nonvanishing  expectation value for a functional of the gauge fields that turns out to be Zwanziger's horizon function. The action allowing to implement this condition is the Gribov-Zwanziger action. This establishes  in a precise way the equivalence between Gribov's no-pole condition and Zwanziger's horizon condition.

\end{abstract}

\baselineskip=13pt


\section{Introduction}

It is well known that the gauge fixing quantization procedure of Yang-Mills gauge theories suffers from ambiguities related to the existence  of the so called Gribov copies \cite{Gribov:1977wm}. For example, zero modes of the Faddeev-Popov operator, which are Gribov copies. These zero modes give rise to singularities in the ghost propagator as the latter is nothing else than the inverse Faddeev-Popov operator. Gribov was the first to point out this problem and to propose a possible resolution \cite{Gribov:1977wm}. In the Landau gauge it consists of restricting the domain of integration in the Euclidean functional integral to the Gribov region $\Omega$, defined as the region in field space where the Faddeev-Popov operator is strictly positive. The region $\Omega$ is known to be convex and bounded in all directions in field space. Moreover, every gauge orbit crosses $\Omega$ at least once\footnote{See \cite{Vandersickel:2012tz} for a recent account on the Gribov problem. Other approaches not directly based on the Gribov region might exist, see for example \cite{Serreau:2012cg}. }.  The boundary $\partial \Omega$ of the region $\Omega$, where the first vanishing eigenvalue of the Faddeev-Popov operator appears, is called the Gribov horizon. Following \cite{Gribov:1977wm}, The implementation of the restriction to $\Omega$ amounts to impose  that the ghost propagator, {\it i.e.}~the inverse of the Faddeev-Popov operator, has no poles at finite non-vanishing values of the momentum $k$. This implies that, within the region $\Omega$, the ghost propagator remains always positive, namely the Gribov horizon $\partial \Omega$ is never crossed.  The only allowed pole is at $k^2=0$, which has the meaning of approaching the horizon $\partial \Omega$. The requirement of absence of poles for the ghost propagator is known as the no-pole condition.

In his seminal work \cite{Gribov:1977wm}, Gribov worked out the no-pole condition at the first non-trivial order and evaluated the ensuing modifications of the gauge and ghost propagators. Subsequently, Zwanziger \cite{Zwanziger:1988jt, Zwanziger:1989mf, Zwanziger:1992qr} provided an independent framework for the restriction to the Gribov region. More precisely, by making use of degenerate quantum mechanics perturbation theory, he has been able to provide a characterization of the eigenvalues $\lambda_n(A)$ of the Faddeev-Popov operator, ${\cal{M}}^{ab} = - ( \partial^2 \delta^{ab }- gf^{abc}A^c_\mu\partial_\mu)$, taking the Laplacian, $-\partial^2 \delta^{ab }$, as starting point  and then performing a resummation of the whole perturbative series. In that way he ended up with a closed expression for the trace of the Faddeev-Popov operator,  $\text{Tr}{\cal{M}}$, which resulted in a nonlocal functional of the gauge field, called the horizon function \cite{Zwanziger:1988jt, Zwanziger:1989mf, Zwanziger:1992qr}. Moreover, relying on the equivalence between the micro-canonical and the canonical ensembles, he constructed a local and renormalizable action implementing the restriction to the region $\Omega$. The ordinary Faddeev-Popov action gets modified by the addition of the horizon function which can be cast in local form by introducing a suitable set of auxiliary fields. The resulting action is known as the Gribov-Zwanziger action.

It is worth underlining that Gribov's no-pole condition and Zwanziger's construction yield exactly the same results at the lowest nontrivial order. For example, the first order gap equation stemming from Gribov's no-pole condition is precisely the same as that obtained through Zwanziger's horizon condition. Also, an explicit two loop computation confirms this \cite{Gracey:2005cx}. Moreover, the lowest order nonlocal modification of the Faddeev-Popov action obtained by Gribov coincides with what one obtains within Zwanziger's construction. This has naturally led to conjecture that both approaches should be equivalent. Though, so far, an all order proof of this statement is still lacking. The aim of the present Letter is to fill this gap.

To some extent, the possible equivalence between the two approaches could also be expected by noticing that the ghost propagator ${\cal{G}}(x,y;A)$ in an external gauge field $A$ can be expressed as
\begin{equation}
{\cal{G}}(x,y;A) = \sum_n \frac{u_n(x;A) u^*_n(y;A) } {\lambda_n(A)}  \;, \label{sum}
\end{equation}
where $u_n(x;A)$ stands for the eigenfunction corresponding to the eigenvalue $\lambda_n(A)$ of the Faddeev-Popov operator. Ones realizes thus that positivity of  $ {\cal{G}}(x,y;A)$ is strictly related to that of the eigenvalues $\lambda_n(A)$.

The connection between Zwanziger's construction of the Gribov-Zwanziger action and the original no-pole condition proposed by Gribov was first discussed in a previous work \cite{Gomez:2009tj} by some of the authors of the present work. In \cite{Gomez:2009tj}, it was shown that Zwanziger's  horizon function can be matched to the ghost form factor, defined through the no-pole condition,  up to the third order in the external gauge field. In this work we generalize this result and present a more precise statement of the aforementioned equivalence. We evaluate Gribov's ghost form factor as an infinite series in the external gauge fields, providing the expression of the generic $n$-th term of the expansion.  Further, we show that, at zero external momentum, the whole series can be resummed, the resulting expression coinciding precisely with Zwanziger's horizon function.  It follows thus that the no-pole condition can be expressed as a condition on the $1PI$ expectation value of the ghost form factor. However, this condition cannot be realized within the Faddeev-Popov functional measure, as also pointed out in \cite{Dudal:2009xh}. Instead, it can be consistently implemented by employing the Gribov-Zwanziger action, and thus corresponds to the Zwanziger's horizon condition. This result confirms and proves in a precise way the conjecture that the Gribov-Zwanziger action corresponds to the restriction to the Gribov region in the way originally intended by Gribov.

This Letter is organized as follows: in Sect.~2  brief reviews of the original Gribov no-pole condition and of the construction of Zwanziger's horizon function are given. In Sect.~3 we construct the exact expression of Gribov's ghost form factor as a function of the external momentum and gauge fields and show that it coincides with  Zwanziger's horizon function. Sect.~4 contains our conclusions.

\section{Gribov's no-pole condition and Zwanziger's horizon function}

In this section we give a brief review of the two original proposals for dealing with the Gribov problem, namely Gribov's no-pole condition and  Zwanziger's construction of the Gribov-Zwanziger action.

\subsection{Gribov's no-pole condition}

Gribov \cite{Gribov:1977wm}  pointed out that  even after imposing the Landau gauge condition,  $\partial_\mu A^{a}_\mu=0$,  there still remain redundant gauge copies\footnote{It is worth to point out that the existence of the Gribov copies is not limited to the Landau gauge. It is a feature of the gauge fixing procedure \cite{Singer:1978dk}, given certain assumptions of Lorentz covariance etc.}. In the path integral quantization procedure, these Gribov copies lead to a breakdown of the Faddeev-Popov prescription resulting in an ill-defined  functional measure. In a very schematic exposition, the Faddeev-Popov procedure relies on expressing a unity as ``$\delta$-function $\times$ Jacobian'' where the $\delta$-function imposes the gauge fixing. The combination with the Jacobian determinant can then be lifted into the action using the Faddeev-Popov ``trick'' (= introduction of ghosts), leaving us with a new gauge fixed partition function.   However, the assertion ``$1=\delta$-function $\times$ Jacobian'' needs to be replaced by ``1=$\sum \delta$-function $\times$ Jacobian'' if the gauge fixing condition has multiple solutions. Unfortunately, that sum cannot be lifted into the partition function and the Faddeev-Popov procedure thus fails.

Examples of Gribov copies  can be easily constructed  by looking at zero modes of the Faddeev-Popov operator. Indeed, if $A_{\mu}$ and $A'_{\mu}$ are connected by an infinitesimal gauge transformation
\begin{align}
A'^a_{\mu} = A^a_{\mu} - D^{ab}_{\mu}\omega^b; \;\;\; \text{where}\; D^{ab}_{\mu} = \partial_{\mu} \delta^{ab} -g f^{abc}A^{c}_{\mu}
\end{align}
and both satisfy the Landau gauge condition $\partial_\mu A^{a}_\mu=\partial_\mu A'^{a}_\mu=0$, it follows that  $\omega^a$ is a zero mode of the Faddeev-Popov operator, {\it i.e.}
\begin{align}
- \partial_\mu D^{ab}_{\mu}\omega^b = {\cal M}^{ab} \omega^b = 0 \label{zm} \;.
\end{align}
In \cite{Gribov:1977wm} some examples of nontrivial solutions of this equation were first provided.  In fact, it can be shown that there are an infinite number of zero modes \cite{Guimaraes:2011sf, Capri:2012ev}. It is therefore clear that the Faddeev-Popov quantization does not provide a well-defined measure of integration over non-Abelian gauge fields.

As already mentioned above, this problem is faced by restricting the domain of integration in the path integral to the region  $\Omega$, defined as the set of field configurations obeying the Landau condition and for which the Faddeev-Popov operator ${\cal M}^{ab}$ is strictly positive, namely
\begin{align}
\Omega \;= \; \{ A^a_{\mu}\;; \;\; \partial_\mu A^a_{\mu}=0\;; \;\; {\cal M}^{ab}=-(\partial^2 \delta^{ab} -g f^{abc}A^{c}_{\mu}\partial_{\mu})\; >0 \; \} \;. \label{gr}
\end{align}
In practice, to restrict the domain of integration to the region $\Omega$, Gribov studied the ghost propagator, which is the inverse of the operator ${\cal M}^{ab}$. He started by writing a general expression for the normalized trace of the ghost propagator as a function of the gauge field configuration $A$ and the external ghost momentum $k$
\begin{align}
 {\cal G}(k,A) = \frac{1}{(N^2-1)}\delta^{ab}({\cal M}^{-1})^{ab}  = \frac{1}{k^2} (1+\sigma(k,A)) \;, \label{gp}
\end{align}
where we have defined the ghost form factor $\sigma(k,A)$ in the presence of an external gauge background, see also \eqref{gpara}.
Being interested in the modifications of the gluon propagator in the deep infrared regime, he focused on the contribution to the ghost form factor $\sigma(k,A)$ coming from quadratic terms in the external gauge fields $A^{a}_\mu$.  Up to this order, for $\sigma(k,A)$ one gets
\begin{equation}
\sigma(k,A) = \frac{g^2 N}{N^2-1} \frac{1}{k^2} \int \frac{d^4q}{(2\pi)^4} \frac{(k-q)_\mu k_\nu}{(k-q)^2} A^a(-q)_\mu A^a_\nu(q)  \; + O(A^3) \;, \label{quad}
\end{equation}
and expression (\ref{gp}) can be written as
\begin{equation}
{\cal G}(k,A) \approx \frac{1}{k^2} \frac{1}{1-\sigma(k,A)} \;. \label{np}
\end{equation}
As $\sigma(k,A)$ turns out to be a decreasing function of the momentum $k$ \cite{Gribov:1977wm}, Gribov required the validity of the condition
\begin{align}
\sigma(0,A) \le 1\;, \label{npc}
\end{align}
which is known as the no-pole condition. From condition (\ref{npc}) it follows that the ghost propagator has no poles at finite non-vanishing values of the momentum $k$. Therefore, expression (\ref{np}) stays always positive, meaning that the Gribov horizon $\partial \Omega$ is never crossed.

In order to proceed further with the construction of a well-defined measure for the gauge path integral, the condition \eqref{npc} has to be incorporated into the dynamics of the theory. To that purpose, Gribov  \cite{Gribov:1977wm} modified the Faddeev-Popov measure by including condition  \eqref{npc} through a unit step function $\theta(x)$, {\it i.e.}
\begin{align}
d\mu_{FP}   & = {\cal D}A\; \delta(\partial A)\; \det({\cal M}^{ab})\; e^{-S_{YM}}  \Rightarrow  {\cal D}A\; \delta(\partial A)\; \det({\cal M}^{ab})\; \theta(1-\sigma(0,A)) \; e^{-S_{YM}}  \;, \label{fpm}
\end{align}
where $S_{YM}$ is the classical Euclidean Yang-Mills action
\begin{equation}
S_{YM} = \frac{1}{4} \int d^Dx \; F^a_{\mu\nu}F^a_{\mu\nu} \;. \label{ym}
\end{equation}
Making use of the integral representation
\begin{equation}
\theta(x) = \int_{-i\infty +\varepsilon}^{+i\infty +\varepsilon} \frac{d\beta}{2\pi i\beta} e^{-\beta x} \;, \label{theta}
\end{equation}
it turns out that the Yang-Mills action gets modified by the addition of the factor $\sigma(0,A)$
\begin{equation}
e^{-S_{YM}} \Rightarrow \; e^{-(S_{YM}+\beta\sigma(0,A))}  \;. \label{ymm}
\end{equation}
Therefore, for the partition function $\cal Z$, one writes
\begin{equation}
{\cal Z} = \int {\cal D}A\; \frac{d\beta}{2\pi i\beta}\; \delta(\partial A)\; \det({\cal M}^{ab})\;  e^{-S_{YM}} \;e^{\beta(1-\sigma(0,A))} \;. \label{pf}
\end{equation}
Following e.g.~\cite{Gribov:1977wm,Vandersickel:2012tz}, the integration over $\beta$ can be done in a saddle point approximation in the thermodynamic limit, yielding
\begin{equation}
{\cal Z} = {\cal N}\int {\cal D}A\; \delta(\partial A)\; \det({\cal M}^{ab})\;  e^{-(S_{YM}+\beta^{*}\sigma(0,A) - \beta^*)} \;, \label{pfsp}
\end{equation}
with $\beta^{*}$ determined by the gap equation \cite{Gribov:1977wm}
\begin{equation}
1= \frac{3Ng^2}{4} \int \frac{d^Dk}{(2\pi)^D} \frac{1}{k^4+\frac{g^2N}{2(N^2-1)}{\beta^{*}}} \;. \label{gge}
\end{equation}
A technical argumentation can be found in \cite{Vandersickel:2012tz}. Let us point out that this procedure is based on the approximate form of eqs.\eqref{quad},(\ref{np}), which is only valid up to second order in the external gauge fields. In Section 3 we shall work out the {\it exact} expression of $\sigma(k,A)$. In this case, the implementation of the no-pole condition demands us to consider the gauge field dynamics (the interaction structure) from the beginning. We will show that the no-pole condition turns out to be a statement about the $1PI$ diagrams of the ghost form factor.

\subsection{Zwanziger's horizon condition}

An independent implementation of the restriction to the region $\Omega$ has been worked out by Zwanziger \cite{Zwanziger:1988jt, Zwanziger:1989mf, Zwanziger:1992qr}, through the analysis of the eigenvalues $\lambda(A)$ of the Faddeev-Popov operator
\begin{align}
{\cal M}^{ab} \omega^b = \lambda(A)\omega^a \label{ev}
\end{align}
The Gribov horizon $\partial \Omega$ can be probed by studying the behavior of the smallest eigenvalue $\lambda_{min}(A)$ as a function of the gauge fields configuration. In terms of $\lambda_{min}(A)$, the restriction to the region $\Omega$ is achieved by demanding that
\begin{align}
\lambda_{min}(A) \ge 0 \label{np2} \;,
\end{align}
which implies that the Faddeev-Popov operator is always positive. Zwanziger was able to find an expression for the trace of the operator $\cal M$:
\begin{align}
\text{Tr}\; {\cal M} =  VD(N^2-1) - H(A)    \;. \label{trace}
\end{align}
where $D$ and $V$ stand for the dimensions and the volume of the Euclidean space-time, respectively, and the horizon function $H(A)$ is given by
\begin{align}
H(A)  =  g^{2}\int d^{4}x\;d^{4}y\; f^{abc}A_{\mu}^{b}(x)\left[ {\cal M}^{-1}\right]^{ad} (x,y)f^{dec}A_{\mu}^{e}(y)   \;. \label{hf}
\end{align}
The expression \eqref{hf} is known as the horizon function. In  \cite{Zwanziger:1989mf} (cf. also a more recent discussion in \cite{Vandersickel:2012tz}), it has been argued that, in the infinite volume limit, the condition \eqref{np2} is very well approximated by the demand that $\text{Tr}\; {\cal M}$ is positive, \eqref{trace}, namely
\begin{align}
VD(N^2-1) - H(A) \ge 0  \;. \label{np3}
\end{align}
Relying on the equivalence between the canonical and microcanonical ensembles in the thermodynamic limit, Zwanziger has been able to implement the constraint in the functional integral \cite{Zwanziger:1988jt, Zwanziger:1989mf, Zwanziger:1992qr}. This has resulted in the following partition function
\begin{align}
Z = \;
\int {\cal D}A\; \delta(\partial A)\; \det({\cal M}^{ab})\;  e^{-(S_{YM}+\gamma^4 H(A) -\gamma^4VD(N^2-1))} \;, \label{zw1}
\end{align}
where the massive parameter $\gamma$ is a dynamical parameter determined in a self-consistent way through the horizon condition \cite{Zwanziger:1988jt, Zwanziger:1989mf, Zwanziger:1992qr}
\begin{equation}
\left\langle H(A)   \right\rangle_{GZ} =VD\left(  N^{2}-1\right) \;. \label{hc}
\end{equation}
where we made explicit the fact that the expectation value $ \left\langle H(A)   \right\rangle_{GZ} $ has to be evaluated with the modified action
\begin{align}
S_{GZ} = S_{FP}+\gamma^4 H(A)  -\gamma^4VD(N^2-1) \;, \label{GZ}
\end{align}
where $S_{FP}$ stands for the Yang-Mills action $S_{YM}$ supplemented with the gauge fixing factors coming from the Faddeev-Popov measure, given in \eqref{zw1}. The action $S_{GZ}$, \eqref{GZ}, is known as the Gribov-Zwanziger action. Eqs. \eqref{zw1},  \eqref{hc} implement the restriction to the Gribov region $\Omega$ within Zwanziger's framework. In particular, the horizon condition  \eqref{hc} enables us to express the parameter $\gamma$ as a function of the gauge coupling. At lowest order, condition (\ref{hc}) reads
\begin{equation}
1= \frac{3Ng^2}{4} \int \frac{d^Dk}{(2\pi)^D} \frac{1}{k^4+2g^2N\gamma^4} \;, \label{zge}
\end{equation}
from which one sees that, apart from a numerical coefficient,  the parameter $\gamma^4$ can be identified with $\beta^{*}$, more precisely $\beta^*=4(N^2-1)\gamma^4$ in Gribov's approach.

From this brief review, it should be apparent that, even though Gribov and Zwanziger follow completely different paths to constrain the gauge measure to the horizon $\Omega$,  at the leading order explicitly analyzed by Gribov, their prescriptions for the gauge quantization coincide\footnote{This equivalence was also established up to the third order in the external gauge fields \cite{Gomez:2009tj}.}. This fact strongly suggests that it should be possible to establish a more precise relation between these approaches. This is exactly the purpose of the next section.

\section{The exact ghost form factor and the horizon function}  \label{gm}
In this section we show by an explicit computation that the exact, all-order, ghost form factor  is proportional to the Horizon function. This will lead to a precise connection between the Gribov's no-pole condition and the Zwanziger's horizon condition discussed in the previous sections.

We start by providing an explicit derivation of the ghost two-point correlator as a function of the external gauge fields and external momenta. Such quantity is defined by the Fourier transform of the expression
\begin{align}
\langle \bar{c}^a(x) c^b(y) \rangle = \frac{\int{\cal D}c{\cal D}\bar{c} \;\bar{c}^a(x) c^b(y) e^{\int \bar{c}^c {\cal M}^{cd} c^d }}{\int{\cal D}c {\cal D} \bar{c}\;e^{\int \bar{c}^c {\cal M}^{cd} c^d}}\label{g2pf}
\end{align}
where $\int \bar{c}^a {\cal M}^{ab} c^b$ stands for
\begin{align}
\int \bar{c}^a {\cal M}^{ab} c^b &= \int d^D x \bar{c}^a(x) \left(-\delta^{ab} \partial^2 -gA^{ab}_{\mu}(x)\partial_{\mu} \right) c^b(x)\nonumber\\
&= \int \frac{d^D p}{(2\pi)^D} \int \frac{d^D q}{(2\pi)^D} \bar{c}^a(-p) \left(q^2\delta^{ab}\delta(p-q) - gA^{ab}_{\mu}(p-q)iq_{\mu} \right) c^b(q)\nonumber\\
&= \int \frac{d^D p}{(2\pi)^D} \int \frac{d^D q}{(2\pi)^D} \bar{c}^a(-p) {\cal M}^{ab}(p-q)  c^b(q)  \label{a1}
\end{align}
with $A^{ab} \equiv f^{acb}A^{c}$. We did not use different symbols for  the Fourier transform in expression \eqref{a1}. It is worth noticing that, so far,  expression \eqref{g2pf} does not require the specification of an action describing the dynamics of the gauge fields. This is an important point since we will eventually impose a condition on the ghost two-point function whose fulfillment will require the employment of a specific action.

The expression \eqref{g2pf} can be explicitly evaluated to all orders  by standard techniques, using the Wick theorem, yielding
\begin{align}
\langle \bar{c}^a(x) c^b(y) \rangle &=  \delta^{ab}{\cal G}_0(x-y) + g\int d^D x_1 \;{\cal G}_0(x-x_1)  A^{ab}_{\mu_1}(x_1) \partial^{x_1}_{\mu_1}  {\cal G}_0(x_1-y)\nonumber\\  &+ g^2 \int d^D x_1 \int d^D x_2 \;{\cal G}_0(x-x_1)  A^{ac}_{\mu_1}(x_1) \partial^{x_1}_{\mu_1}  {\cal G}_0(x_1-x_2) A^{cb}_{\mu_2}(x_2) \partial^{x_2}_{\mu_2}  {\cal G}_0(x_2-y) \nonumber\\
&+ \cdots \label{g2pf-2}
\end{align}
where the  expression of order $n$ in $g$ has the form
\begin{align}
  g^n \int d^D x_1 \int d^D x_2 \cdots \int d^D x_n \; &\left[ {\cal G}_0(x-x_1)  A^{aa_1}_{\mu_1}(x_1) \partial^{x_1}_{\mu_1}  {\cal G}_0(x_1-x_2) A^{a_1a_2}_{\mu_2}(x_2) \partial^{x_2}_{\mu_2}  {\cal G}_0(x_2-y)\right.\nonumber\\
 &\left. \cdots A^{a_{n-1}b}_{\mu_n}(x_n) \partial^{x_n}_{\mu_n}  {\cal G}_0(x_n-y)\right]
\end{align}
whereby  ${\cal G}_0(x-y)$ stands for the free  ghost propagator
\begin{align}
 {\cal G}_0(x-y) =  \int \frac{d^D p}{(2\pi)^D}  \frac{1}{p^2} e^{ip(x-y)}\,.
 \end{align}
It is useful to rewrite eq.(\ref{g2pf-2}) in Fourier space
\begin{align}
\langle \bar{c}^a(p) c^b(-q) \rangle &= \int d^D x \; \int d^D y \langle \bar{c}^a(x)  c^b(y)\rangle e^{-ipx}e^{iqy}\nonumber\\
&= \frac{1}{p^2} \left[\delta^{ab}\delta(p-q) + g A^{ab}_{\mu}(p-q) \frac{iq_{\mu}}{q^2}   \right.\nonumber\\
&+g^2 \int \frac{d^D r}{(2\pi)^D}  A^{ac}_{\mu}(p-r) \frac{ir_{\mu}}{r^2} A^{cb}_{\nu}(r-q) \frac{iq_{\nu}}{q^2} +\cdots \nonumber\\
&\left.+g^n \int \frac{d^D q_1}{(2\pi)^D} \cdots \int \frac{d^D q_{n-1}}{(2\pi)^D}  A^{aa_1}_{\mu_1}(p-q_1) \frac{iq_{1\mu_1}}{q_1^2} A^{a_1a_2}_{\mu_2}(q_1-q_2) \frac{iq_{2\mu_2}}{q_2^2}  \cdots A^{a_{n-1}b}_{\mu_n}(q_{n-1}-q) \frac{iq_{\mu_n}}{q^2} + \cdots \right]\,. \label{g2pf-3}
\end{align}
Following Gribov  \cite{Gribov:1977wm}, we look at the full normalized trace of expression  \eqref{g2pf-3}, that is
\begin{align}
{\cal G} (k,A) &= \frac{1}{V(N^2-1)} \langle \bar{c}^a(p) c^a(-q) \rangle\vert_{p=q=k}= \frac{1}{k^2} (1 + \sigma(k,A)) \label{gpara}
\end{align}
where $V = \delta(p-q)\vert_{p=q}$. It then follows from eq.(\ref{g2pf-3}) that the exact ghost form factor is given by
\begin{align}
\sigma (k,A) &= \frac{1}{V(N^2-1)}\left[g^2 \int \frac{d^D r}{(2\pi)^D}  A^{ac}_{\mu}(k-r) \frac{ir_{\mu}}{r^2} A^{ca}_{\nu}(r-k) \frac{ik_{\nu}}{k^2} +\cdots \right.\nonumber\\
&\left.+g^n \int \frac{d^D q_1}{(2\pi)^D} \cdots \int \frac{d^D q_{n-1}}{(2\pi)^D}  A^{aa_1}_{\mu_1}(k-q_1) \frac{iq_{1\mu_1}}{q_1^2} A^{a_1a_2}_{\mu_2}(q_1-q_2) \frac{iq_{2\mu_2}}{q_2^2} \cdots  A^{a_{n-1}a}_{\mu_n}(q_{n-1}-k) \frac{ik_{\mu_n}}{k^2} + \cdots \right]\,.
\end{align}
Let us now show that, in the limit of vanishing external momentum $k\rightarrow 0$,  the ghost form factor $\sigma(k\rightarrow 0, A)$ is proportional to Zwanziger's horizon function. Focusing on the general term of order $n$ in $g$, we perform the change of variables $q_i \rightarrow k + q_i$, for $i=1,\cdots,n-1$, obtaining
\begin{align}
&g^n \int \frac{d^D q_1}{(2\pi)^D} \cdots \int \frac{d^D q_{n-1}}{(2\pi)^D}  A^{aa_1}_{\mu_1}(-q_1) \frac{ik_{\mu_1}}{(k+q_1)^2} A^{a_1a_2}_{\mu_2}(q_1-q_2) \frac{i(k+q_2)_{\mu_2}}{(k+q_2)^2}
 \cdots A^{a_{n-1}a}_{\mu_n}(q_{n-1}) \frac{ik_{\mu_n}}{k^2}= -\frac{k_{\mu_1}k_{\mu_n}}{k^2} f^{(n)}_{\mu_1\mu_n}
\end{align}
where the transversality of $A^{ab}_{\mu}$ in the Landau gauge was used. The tensor $ f_{\mu_1\mu_n}$ depends on the momentum $k$ and on the gauge field polarizations. Therefore, we conclude that\footnote{We notice that, being a function of the momentum $k_\mu$ and of the external gauge field $A^a_\mu$, the tensor $ f_{\mu \nu}$ can be expressed in the following general form
\begin{equation}
 f_{\mu \nu} = \alpha_1(k;A) \delta_{\mu\nu} + \alpha^{ab}_2(k;A)A^a_\mu A^b_ \nu + \alpha_3^{a}(k;A) k_\mu A^a_\nu + \alpha_4^a(k;A) k_\nu A_\mu^a + \alpha_5(k;A) k_\mu k_\nu  \;, \label{decf}
 \end{equation}
 where $\alpha_i, \; i=1,...,5$ are scalar quantities. Thus, equation \eqref{limit} follows due to the transversality of the gauge field, $k_\mu A^a_\mu=0$. }
\begin{align}
 \lim_{k\rightarrow 0} -\frac{k_{\mu_1}k_{\mu_n}}{k^2} f^{(n)}_{\mu_1\mu_n} = -\frac{g^n}{D} \int \frac{d^D q_1}{(2\pi)^D} \cdots \int \frac{d^D q_{n-1}}{(2\pi)^D}  A^{aa_1}_{\mu}(-q_1) \frac{1}{q_1^2} A^{a_1a_2}_{\mu_2}(q_1-q_2) \frac{iq_{2\mu_2}}{q_2^2}
 \cdots A^{a_{n-1}a}_{\mu}(q_{n-1})\,.  \label{limit}
\end{align}

In order to obtain a closed expression for $\sigma(0, A)$, it is useful to introduce a matrix notation. Defining
\begin{align}
\mathds{A}^{ab}_{pq}  = A^{ab}_{\mu}(p-q) \frac{iq_\mu}{q^2}
\end{align}
with matrix multiplication defined by
\begin{align}
(\mathds{A}^2)^{ab}_{pq}  = \int \frac{d^D r}{(2\pi)^D}  A^{ac}_{\mu}(p-r) \frac{ir_{\mu}}{r^2} A^{cb}_{\nu}(r-q) \frac{iq_{\nu}}{q^2}
\end{align}
we can write the general term of order $n$ in $g$ as
\begin{align}
-\frac{g^n}{D} \int \frac{d^D q_1}{(2\pi)^D} \int \frac{d^D q_{n-1}}{(2\pi)^D}  A^{aa_1}_{\mu}(-q_1) \frac{1}{q_1^2} (\mathbb{A}^{n-2})^{a_1a_{n-1}}_{q_1q_{n-1}}A^{a_{n-1}a}_{\mu}(q_{n-1})  \;.
\end{align}
Thus
\begin{align}
\sigma(0, A) =   -\frac{g^2}{VD(N^2-1)} \int \frac{d^D p}{(2\pi)^D} \int \frac{d^D q}{(2\pi)^D}  A^{ab}_{\mu}(-p) \frac{1}{p^2} \left(\sum_{n=0}^{\infty} (g\mathds{A})^{n}\right)^{bc}_{pq}A^{ca}_{\mu}(q)
\end{align}
Now, in matrix notation, we also have
\begin{align}
{\cal M}^{ab}(p-q) &=  q^2\delta^{ab}\delta(p-q) - gA^{ab}_{\mu}(p-q)iq_{\mu}= q^2 (\mathds{1} - g\mathds{A})^{ab}_{pq}
\end{align}
where $\mathds{1} = \delta^{ab}\delta(p-q)$. Then, the  ghost propagator can be written as
\begin{align}
{\cal M}^{-1} =  \frac{1}{p^2} \left[\mathds{1} - g\mathds{A}\right]^{-1}= \frac{1}{p^2}\sum_{n=0}^{\infty} (g\mathds{A})^{n}\,,
\end{align}
from which we finally obtain the exact expression for the ghost form factor
\begin{align}
\sigma(0, A) &=   -\frac{g^2}{VD(N^2-1)} \int \frac{d^D p}{(2\pi)^D} \int \frac{d^D q}{(2\pi)^D}  A^{ab}_{\mu}(-p) \left({\cal M}^{-1}\right)^{bc}_{pq}A^{ca}_{\mu}(q)=\frac{H(A)}{DV(N^2-1)}\,. \label{ff-hf}
\end{align}
Finally, as anticipated above, one sees that the exact expression for the ghost form factor at zero momentum is directly proportional to the horizon function. This will provide us with the precise connection between the original Gribov no-pole condition and  the Zwanziger construction of the Gribov-Zwanziger action.

In order to establish this connection we first note that, beyond first order, {\it i.e.}~the quadratic approximation, we cannot straightforwardly express eq.(\ref{gpara}) as in eq.(\ref{np}), with the gauge field as an external field. The same is in general true also for the expectation values, which reflects the fact that the gauge field dynamics must be considered from the beginning. As is well established from general properties of quantum field theory, the precise statement has to be done in terms of $1PI$ diagrams, namely
\begin{align}
{\cal G} (k) = \langle {\cal G} (k,A)\rangle^{conn} = \frac{1}{k^2} (1 + \langle\sigma(k,A)\rangle^{conn}) =  \frac{1}{k^2} \frac{1}{(1 - \langle\sigma(k,A)\rangle^{1PI})}\label{np-gz}
\end{align}
where ``$conn$'' stands for the connected set of diagrams and $1PI$ denotes the one-particle irreducible ones\footnote{Notice that in eq.\eqref{np-gz}, the gauge field $A$ is no more an external field since, although not yet specified, the expectation value $\langle\sigma(k,A)\rangle$ is meant to be evaluated with an appropriate functional measure allowing to impose the horizon condition \eqref{np-gz-cond-3}.}. From this expression, the no-pole condition for the  ghost form factor reads
\begin{align}
\langle\sigma(0,A)\rangle^{1PI} \le 1 \label{np-gz-cond}
\end{align}
From equation eq.(\ref{ff-hf}), we can see that this condition is equivalent to
\begin{align}
 VD(N^2-1) - \langle H(A)\rangle^{1PI} \ge 0\,, \label{np-gz-cond-2}
\end{align}
which represents a no-pole condition valid to all orders in the gauge coupling.

Notice that we have not yet specified  the dynamics of the gauge field, that is, we have not yet defined an action for the gauge fields with which the expectation value in eq.(\ref{np-gz-cond-2}) is to be computed. Actually, our task of defining a gauge path integral that obeys Gribov's no-pole condition at all orders becomes exactly that of finding a gauge action which is capable of implementing the inequality in eq.(\ref{np-gz-cond-2}).

From the previous section, we know that the Gribov-Zwanziger action provides a framework compatible with the horizon condition, which in the infinite volume limit takes the form
\begin{align}
\langle H(A)\rangle^{1PI}_{GZ} =  VD(N^2-1)
\,. \label{np-gz-cond-3}
\end{align}
This result also reveals the interesting fact that the horizon condition in Gribov-Zwanziger formulation is made up only by diagrams which are $1PI$. Observe that the horizon condition, eq.(\ref{hc}),  can be written as a stationary condition for the vacuum energy:
\begin{align}
\frac{\partial {\cal E}}{\partial \gamma^2} = 0  \label{stat}
\end{align}
where the vacuum energy ${\cal E}$ is defined from
\begin{align}
e^{-{\cal E}} = \;
\int {\cal D}A\; \delta(\partial A)\; \det({\cal M}^{ab})\;  e^{-(S_{YM}+\gamma^4 H(A) - \gamma^4VD(N^2-1))} \;. \label{vac}
\end{align}
Eq. \eqref{stat} is easily seen to give  $\langle H(A)\rangle^{conn}_{GZ} =  VD(N^2-1)$.
Nevertheless,  in the Landau gauge, it follows that the only diagrams contributing to the vacuum energy are\footnote{The $1PI$ nature of the horizon condition is directly connected to the fact that it is a condition on a vacuum energy. Diagrammatically, the quantity ${\cal E}$ is composed of bubble contributions, without external legs and, consequently, with no external momentum flow. It is straightforward to see then that any one-particle reducible diagram contributing to ${\cal E}$ is actually proportional to the square of the zero-momentum expectation value of the fundamental field propagating in the reducible line, i.e.~to the condensate of this fundamental field. Due to Lorentz and/or global color invariance, these condensates are forbidden and therefore all $1PR$ contributions to the vacuum energy vanish.} $1PI$. As a consequence,  $\langle H(A)\rangle^{conn}_{GZ} =  \langle H(A)\rangle^{1PI}_{GZ}$. It then follows that the gap equation is indeed given by eq.(\ref{np-gz-cond-3}).

We also point out that the infrared limit of the ghost form factor is subtle: the GZ connected average and the $k\to 0$ limit are not trivially interchangeable. At lowest order in the external momentum $k$, one has
\begin{align}
 \langle\sigma(k,A)\rangle^{1PI}_{k\approx 0} \approx \langle\sigma(0,A)\rangle^{1PI} -ck^2
\end{align}
where $c$ is a numerical constant and we have omitted the $GZ$ subscript of the averages for notational simplicity. This leads consistently to the well-known enhanced behavior of the ghost propagator in the deep infrared limit, $k\rightarrow 0$, in the Gribov-Zwanziger framework
\begin{align}
{\cal G} (k)_{k \approx 0} \approx \frac{1}{k^4}\,.
\end{align}
It is worthwhile underlining that an alternative argument behind the $1PI$-nature of the horizon condition and the ensuing infrared enhancement of the ghost was presented in \cite{Zwanziger:1992qr}.

It should be, however, remembered here that, according to the most recent lattice data \cite{Cucchieri:2007md, Sternbeck:2007ug, Cucchieri:2007rg, Cucchieri:2008fc,Maas:2008ri, Bogolubsky:2009dc,  Dudal:2010tf, Cucchieri:2011ig,Oliveira:2012eh,Sternbeck:2012mf}, the ghost propagator is not enhanced in the infrared, having instead the following asymptotic behavior
\begin{align}
{\cal G} (k)_{k \approx 0} \approx \frac{1}{k^2} \label{dec}
\end{align}
while the gluon propagator turns out to be suppressed in the infrared, violating positivity and attaining  a non vanishing value at $k=0$.
In \cite{Dudal:2007cw,Dudal:2008sp,Dudal:2011gd}, a theoretical framework was presented in order to accommodate these results within the Gribov-Zwanziger framework\footnote{A short selection of other approaches is, for example, \cite{Aguilar:2008xm,Boucaud:2008ky,Fischer:2008uz,Tissier:2011ey,Weber:2011nw,Frasca:2007uz,Pennington:2011xs,Dudal:2012zx,Huber:2012kd}.}. This has led to what is now called the Refined Gribov-Zwanziger ($RGZ$) action, whose construction relies on the observation that, besides the parameter $\gamma^2$, additional nonperturbative effects related to dimension two condensates have to be taken into account in the infrared. Remarkably, these condensates can be taken into account by maintaining the renormalizability of the theory, resulting in the so called $RGZ$ action. In view of our previous discussion, it is natural to ask how the $RGZ$ action fits in our general considerations? To that end, we recall that tree level part of the $RGZ$ action can be written as
\begin{align}
S_{RGZ} = S_{GZ} + \frac 12 m^2\int d^Dx A^a_{\mu}A^a_{\mu}  - M^2 \int d^Dx (\bar{\phi}^a_i\phi^a_i - \bar{\omega}^a_i\omega^a_i) + \text{rest}  \;, \label{RGZ}
\end{align}
where $\bar{\phi}^a_i, \phi^a_i,\bar{\omega}^a_i, \omega^a_i$ are the usual auxiliary fields introduced in order to localize the horizon function. $m^2$ and $M^2$ are the vacuum expectation values of $d=2$ scalar fields that condense and whose {\it vev} are in a 1-1 correspondence with the composite operators $A^2$ and $\bar{\phi}\phi -\bar{\omega}\omega$ \cite{Dudal:2011gd}. With this action, it can be easily inferred that the Faddeev-Popov operator appearing in the horizon function will be dynamically transformed into $-\p D+M^2$, i.e.~we loose the direct connection with the inverse ghost form factor, viz.~the quantity $\sigma(0)$. In addition, the vacuum energy in the $RGZ$ setting of \cite{Dudal:2011gd} is also no more of a pure $1PI$ nature due to the presence of condensing scalar fields ($\sim$ dimension two operators). So, even though the expression \eqref{ff-hf} remains valid, we now have that $\langle \sigma(0,A)\rangle^{1PI}_{RGZ} \neq 1$, implying that, within the $RGZ$ framework \cite{Dudal:2008sp}, the ghost is no more enhanced, being in agreement with the lattice results, in eq.\eqref{dec}.

\section{Conclusion}
By expressing the ghost propagator exactly to all orders in the external gauge fields, we have precisely established the long suspected equivalence between Gribov's no pole condition and the GZ scenario. This relies on the crucial technical result that the exact ghost form factor $\sigma(k,A)$ becomes proportional to Zwanziger's horizon function in the infrared limit, $k\to 0$, which guarantees that Gribov's condition of absence of poles in the ghost propagator ultimately translates into a condition for Zwanziger's horizon function.

In a work in progress partially based on the analysis in this Letter, a renormalizable continuum version of the so-called Landau $B$-gauges \cite{Maas:2009se,Sternbeck:2012mf} will be studied, wherein the ghost form factor at zero momentum is introduced as a boundary condition. In our language, this amounts to set $\sigma(0)=B$ with $B\leq 1$ a kind of ``gauge choice''. A thorough analysis of the implementation and ramifications of this will be presented elsewhere.

\section*{Acknowledgments}
This work was partially supported by CNPq.
L.F.P. acknowledges the support of the Alexander von Humbolt Foundation. D.D. thanks the Research-Foundation Flanders. M.S.G. gratefully
acknowledges the hospitality at and support from the UGent where part of this work was executed.

\end{document}